\documentclass[aps,prl,twocolumn,amsmath,amssymb,superscriptaddress,reprint,nobibnotes]{revtex4-2}

\usepackage{amsmath,mathtools}
\usepackage{graphicx}
\usepackage[dvipsnames]{color}
\usepackage{hyperref}
\usepackage{psfrag}
\usepackage{stmaryrd}
\usepackage{amssymb}
\usepackage{wasysym}
\usepackage{float}
\usepackage{xcolor}
\usepackage{centernot}
\usepackage[T1]{fontenc}
\usepackage[french,english]{babel}
\usepackage[utf8]{inputenc}
\usepackage{lmodern}
\usepackage[normalem]{ulem}

\begin{document}

\title{Visualization of Inertial and Kelvin Waves on the Quantum Vortex Lattice in Superfluid Helium}

\author{Florian Lorin}
\affiliation{Univ. Grenoble Alpes, CNRS, Grenoble INP, Institut Néel, 38000 Grenoble, France}
\author{Charles Peretti}
\altaffiliation[Present address: ]{National High Magnetic Field Laboratory, 1800 East Paul Dirac Drive, Tallahassee, Florida 32310, USA}
\affiliation{Univ. Grenoble Alpes, CNRS, Grenoble INP, Institut Néel, 38000 Grenoble, France}
\author{Corentin Bourjaillat}
\affiliation{Univ. Grenoble Alpes, CNRS, Grenoble INP, Institut Néel, 38000 Grenoble, France}
\author{Patrik \v{S}van\v{c}ara}
\affiliation{Univ. Grenoble Alpes, CNRS, Grenoble INP, Institut Néel, 38000 Grenoble, France}
\author{Pierre-Philippe~Cortet}
\affiliation{Universit\'e Paris-Saclay, CNRS, FAST, 91405 Orsay, France}
\author{Mathieu Gibert}
\email[]{mathieu.gibert@neel.cnrs.fr}
\affiliation{Univ. Grenoble Alpes, CNRS, Grenoble INP, Institut Néel, 38000 Grenoble, France}

\date{\today}

\begin{abstract}
Superfluid \textsuperscript{4}He subjected to steady rotation develops a regular lattice of quantum vortices aligned with the rotation axis. We prepare this lattice in a rotating cryostat, perturb it with a constant heat flux, and visualize vortex deformation waves that propagate in the lattice and grow in energy with the forcing. Below twice the rotation rate, we show that these waves feature a continuous frequency spectrum whose structure corresponds to inertial waves. At larger frequencies, we report several observations suggesting the presence of Kelvin waves over a continuous range of frequency. These results, obtained through space- and time-resolved velocity measurements of quantum vortex lines, appear as quantitative constraints for modern two-way coupled vortex filament models. They further offer new insights into perturbed quantum vortex systems in three-dimensional quantum fluids.
\end{abstract}

\maketitle

\textit{Introduction.---} A hallmark property of superfluids is the localization of vorticity on filamentary structures known as quantum vortices~\cite{Barenghi2023,Tsubota2025}. Unlike their classical counterparts, quantum vortices appear as irrotational whirls whose circulation is given by the elementary quantum $\kappa = h/m$, where $h$ is the Planck constant and $m$ is the mass of the bosons constituting the superfluid. Despite their discrete nature, systems of quantum vortices tend to reproduce the behavior of classical fluids at length scales larger than the typical intervortex distance, while nonclassical effects usually emerge at smaller scales~\cite{Barenghi2023,Tsubota2025,LaMantia2016a,Salort2021,Diribarne2021}.

In this Letter, we consider quantum vortices in He~II, i.e. the superfluid phase of \textsuperscript{4}He that occurs below the critical temperature $T_\lambda \simeq 2.177$~K at the saturated vapor pressure. The hydrodynamics of He~II is usually modeled (for temperatures typically above $1$~K) as a spatial superposition of two coupled fluids~\cite{Donnelly2005}: the normal component (a classical viscous fluid) and the inviscid superfluid component that hosts quantum vortices. The relative proportion of these components varies with temperature~\cite{Donnelly1998}. At absolute zero, the density of the normal component vanishes, and He~II can be treated as an ensemble of quantum vortices. Conversely, at higher temperatures, quantum vortices interact with the normal component through the so-called mutual friction, resulting in a coupled dynamics that remains largely to be understood~\cite{Boue2015,Kobayashi2024}.

A remarkable configuration allowing experimental insights into the intricate dynamics of quantum vortices in He~II occurs in a container subjected to steady rotation. In this situation, the superfluid approaches solid-body rotation with the same angular velocity $\Omega$ as the container by creating a regular lattice of rectilinear quantum vortices aligned with the axis of rotation. The distance between the vortices in the lattice, which is steady in the rotating frame, settles at $\delta=\sqrt{\kappa/2\Omega}$, ensuring that the coarse-grained vorticity in the non-rotating frame equals $2\Omega$. This self-organization mechanism was first predicted by Feynman in the 1950s~\cite{Feynman1957}, a decade before the triangular lattice was identified as the lowest-energy configuration~\cite{Tkachenko1966} (see also \cite{Sandier2012,Roman2025}). Experimentally, following the direct visualization of quantum vortices using solid hydrogen particles in 2006~\cite{Bewley2006}, a quantitative confirmation of the ordered vortex lattice in the bulk of rotating He~II has been achieved in 2023~\cite{Peretti2023}.

In the 1980s, Swanson \textit{et al.}~\cite{Swanson1983} studied the stability of the vortex lattice in rotating He~II against perturbations induced by a heat flux oriented along the axis of rotation. Using measurements of second sound attenuation~\cite{Varga2019,Woillez2023}, they estimated the (space-time) average density of quantum vortices per unit volume, and observed two transitions as the heat flux increased from zero. This suggested a stable lattice configuration occurring only at low heat flux, followed by two perturbed regimes. The second transition (at large heating power) was associated with the destruction of the vortex lattice and the onset of quantum turbulence~\cite{Barenghi2023,Tsubota2025}. The first transition was instead suggested to relate to the excitation of waves on quantum vortices, driven by the interaction with the normal fluid component through the so-called Donnelly-Glaberson (DG) instability~\cite{Glaberson1974}.

The deformability of quantum vortices indeed allows multiple wave modes to propagate through the vortex lattice~\cite{Sonin1987} (see Supplemental Material (SM) \cite{SM} for a detailed description). A driven buildup and decay of perturbations of the vortex lattice---potentially corresponding to these wave modes---have been reported in 2023 by Mäkinen \textit{et al.}~\cite{Makinen2023} in rotating superfluid $^3$He, by harmonically modulating the angular velocity of the container. These were revealed by NMR techniques measuring global quantities such as the volume density and average tilt of quantum vortices. Conversely, Minowa \textit{et al.}~\cite{Minowa2025} observed forced deformation waves on an individual quantum vortex in non-rotating He~II by exciting charged nanoparticles, trapped along the vortex core, by oscillating the surrounding electric field.

In this Letter, we investigate the weakly perturbed regime of the vortex lattice in rotating He~II submitted to a heat flux below the transition to a turbulent regime. While the initial arrangement of quantum vortices is tightly controlled by global rotation, we impose a constant heat flux along the rotation axis and visualize the resulting motion of quantum vortices. We discover that, at low heating power, the vortex lattice first hosts an ensemble of propagating inertial waves before a transition to a state composed of inertia-Kelvin waves takes over beyond the threshold of the DG instability. We identify the physical nature of these quantum vortex excitations by measuring the (space-time resolved) velocity field corresponding to the vortex displacements which also allows us to infer how their amplitude scales with the applied heat flux. Our results challenge the previous understanding, based on indirect space-time averaged measurements, of the vortex lattice dynamics.

\textit{Experiments and PIV measurements.---} 
Steady global rotation of He~II around the vertical axis at a rate $\Omega=1.047~\mathrm{rad\,s^{-1}}$ ($10$~rpm) is achieved with \mbox{CryoLEM}, a custom-built helium cryostat mounted on a 1.2~m diameter rotating platform (see \cite{Vessaire2026} for details). The temperature of the liquid helium bath inside the cryostat is regulated at $T = (2.088 \pm 0.005)$~K by controlling the saturated vapor pressure above it via a rotary pumping feedthrough. Under these conditions, a stable lattice of quantum vortices is reproducibly observed in the rotating frame~\cite{Peretti2023}, with the intervortex distance matching the theoretically expected value $\delta = \sqrt{\kappa/2\Omega} \simeq 220~\mu$m, with $\kappa = h/m\simeq 9.97\times10^{-8}~\mathrm{m^2\,s^{-1}}$, where $m$ is the mass of the \textsuperscript{4}He atom.

The vortex lattice is investigated inside a transparent Plexiglas channel of $8$~cm height and $2$~cm horizontal square base, corresponding to a cross section $A = 4 \times 10^{-4}~$m$^2$. The channel is horizontally centered in the cryostat, aligned with its axis of rotation, and fully submerged in the liquid helium bath. Drawing inspiration from Swanson \textit{et al.}~\cite{Swanson1983}, we impose a steady heat flux $\dot{q} = P/A$ along the channel by a planar heater that closes the channel at its bottom end and covers its entire cross section. The heating power $P$ drives a thermal counterflow, a non-classical flow configuration unique to superfluids (see \cite{Skrbek2021} for a comprehensive overview).

The CryoLEM's experimental volume is optically accessible from the side through four sets of windows, aligned with the vertical walls of the experimental channel. To visualize the motion of quantum vortices \cite{Guo2014}, we seed the fluid with micrometric particles of solid H\textsubscript{2}, following a protocol described in \cite{Vessaire2026}. The particles gather at the cores of quantum vortices and therefore allow their direct imaging~\cite{Bewley2006,Peretti2023}. This is achieved by illuminating the particles with an approximately $100~\mu$m thick vertical laser sheet passing through the middle of the experimental cell and capturing their motion under normal incidence by a digital camera mounted on the rotating platform just like the laser.

In order to quantify the motion of the quantum vortices, we process the image time series using a particle image velocimetry (PIV) protocol: we compute cross-correlations between pairs of images, separated by $0.1$~s time intervals, over square windows of $64$~pixel ($\simeq 428~\mu$m) side with $50\%$~overlap. This produces two-dimensional velocity fields ($u_x$,$u_z$) in the ($x$,$z$) vertical plane in the rotating frame with a resolution of $214~\mu$m over a region of $14.5$~mm horizontal and $17.2$~mm vertical extension at the horizontal center of the experimental channel. The bottom end of the field of view lies about $4$~mm above the heater. The time series begin when the channel heater is switched on ($t=0$), and span between 21 and 50 rotation periods of the cryostat. The calculated velocity field reflects the displacements of the quantum vortex cores at a scale slightly coarse-grained relative to the intervortex distance $\delta \simeq 220~\mu$m. This naturally follows from the fact that the majority of particles are trapped in these cores (see the movie in SM).

\textit{Results.---} We report on 22 experiments discerned by different heating powers $P$ ranging from $0$ to $50$~mW, at a fixed rotation rate $\Omega$ and temperature $T$. Beyond the value $P=50$~mW, which numerically corresponds to the second threshold identified by Swanson \textit{et al.} \cite{Swanson1983} (see SM), we observe the vortex lattice to breakup and the flow transition into a turbulent regime, in agreement with previous interpretations based on measurements of the average volume density of quantum vortices~\cite{Swanson1983,Dwivedi2024,Novotny2026}.

\begin{figure}[tbp!]
    \centering
    \includegraphics{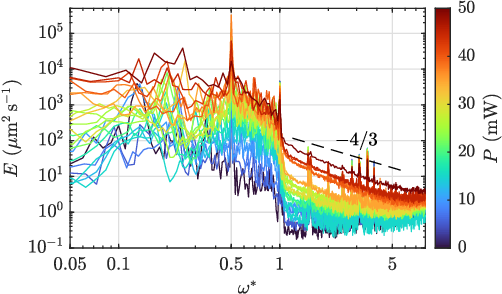}
    \caption{Temporal PSD $E$ of the PIV-computed velocity field as a function of non-dimensional frequency $\omega^*=\omega/2\Omega$ for the series of experiments at heating power $P$ increasing from $0$ to $50$~mW.}
    \label{fig:PSD}
\end{figure}

We begin our analysis by calculating the temporal power spectral densities (PSD) $E_x$ and $E_z$ from the PIV-computed velocity fields $u_x$ and $u_z$. We first compute these spectra for each PIV interrogation window, by taking the (properly normalized) square of the modulus of the temporal Fourier transform of the signal. We then evaluate, at each frequency, the median value across all these windows, and report in Fig.~\ref{fig:PSD} the resulting PSD $E = E_x + E_z$ as a function of the non-dimensional angular frequency $\omega^* = \omega/2\Omega$. A global increase of the spectral energy content with the applied heating power $P$ is readily evident, which indicates a link between the observed response of the vortex lattice and the imposed perturbation. This trend will be quantitatively discussed later, as we will first address several non-trivial features that appear in the spectra.

At low $P$, the spectra are dominated by sharp peaks at $\omega^*=0.5$ and $1$ (corresponding to $\omega=\Omega$ and $2\Omega$), with weaker peaks at their harmonic and nonlinear combinations. By band-pass filtering around these frequencies (see Fig.~2 in SM), we find oscillations propagating in neither the $x$ nor $z$ directions. These are compatible with standing large-scale modes, suggesting a global harmonic perturbation that may arise, e.g., from the sloshing due to platform balancing imperfections (discussed in~\cite{Vessaire2026}) or Earth's rotation, both known to excite such kind of modes in classical rotating fluid systems~\cite{LeBars2015,Triana2012,Boisson2012}.

\begin{figure}[tbp!]
    \centering
    \includegraphics{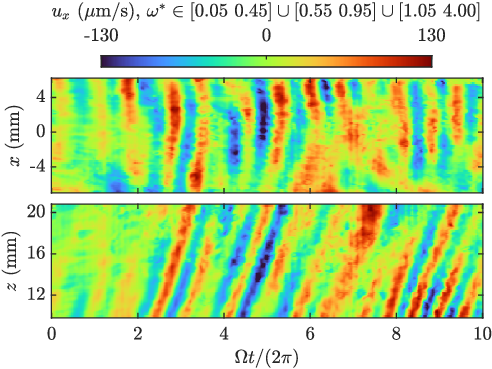}
    \caption{Space-time diagrams of the horizontal velocity $u_x$ for an experiment at $P = 45$~mW, frequency-filtered in the range specified above the figure, for $z = 15$~mm (top panel) and $x = 0$~mm (bottom panel). Time $t$ is normalized by the cryostat's period of rotation $2\pi/\Omega$, and $t=0$ indicates the moment when the heater is switched on. $x=0$ marks the horizontal center of the He~II channel and $z=0$ the vertical position of the heater surface.}
    \label{fig:spacetime}
\end{figure}

Leaving these resonant contributions aside, the spectra suggest that the imposed heat flux excites waves in a broad frequency range. We inspect these excitations in Fig.~\ref{fig:spacetime}, where we display appropriately band-pass filtered space-time diagrams of the horizontal velocity $u_x$ specifically for the experiment at large heating power $P=45$~mW (below the turbulent threshold). We notice that energetic wave modes with frequencies in the range $\omega^* < 1$ emerge soon after the heater activation at $t=0$ and reach a steady amplitude within a few platform rotations. These waves are primarily standing in the $x$ direction (top panel), while slanted structures in the $z$ direction (bottom panel) indicate propagation along the direction of the applied heat flux. Although the diagrams suggest a dominant frequency, a closer look at Fig.~\ref{fig:spacetime} reveals that the dominant mode is subject to frequency fluctuations, eventually yielding the broad frequency spectrum reported in Fig.~\ref{fig:PSD} at $\omega^* < 1$. Higher frequency modes ($\omega^* > 1$) manifest as small-scale ($\sim 500$~$\mu$m) perturbations to the large-scale wave fronts. This unambiguous characterization of the weakly perturbed regime of heated rotating He~II as a collection of propagating waves, impossible from the indirect measurements of Swanson \textit{et al.}~\cite{Swanson1983}, has here been made possible thanks to our direct measurements of the vortex line velocity field.

To better understand the nature of these waves, we return to Fig.~\ref{fig:PSD} and notice a sudden drop in spectral power just beyond the $\omega^* = 1$ peak. We suspect this behavior relates to an interplay between waves propagating on the vortex lattice and in the rotating normal component. The latter indeed supports classical inertial waves (i.e., the helical waves induced by the Coriolis force) whose dispersion relation reads $\omega^* = |k_z|/k$, where $k_z$ is the component along the rotation axis of wavevector ${\bf k}$ of norm $k=|{\bf k}|$~\cite{Mora2021}. These modes therefore propagate only for $\omega^*\leq 1$ and become evanescent at higher frequencies. Thus, the sharp drop in Fig.~\ref{fig:PSD} indicates a potential coupling between wave modes in the normal and superfluid components below $\omega^* = 1$.

\begin{figure}[tbp!]
    \centering
    \includegraphics[width=8.5cm]{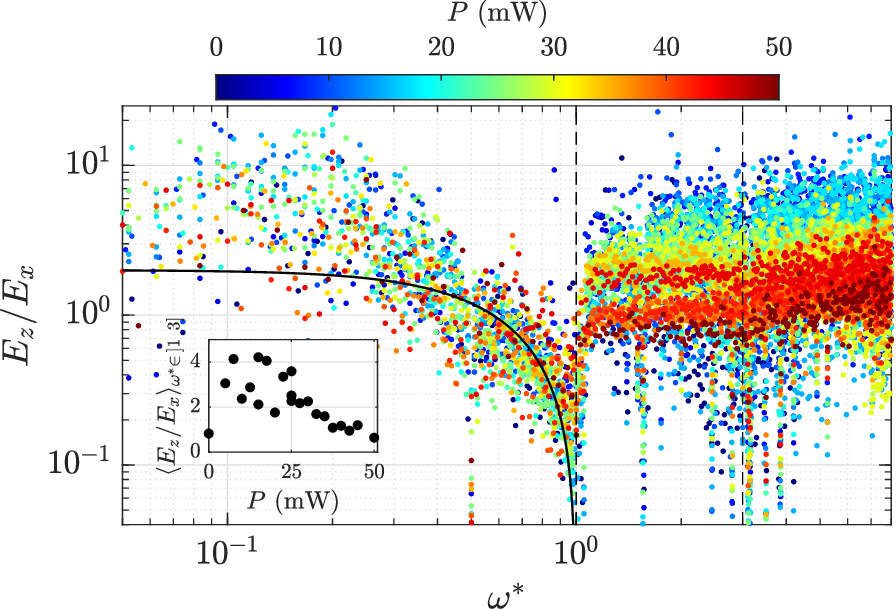}
    \caption{Ratio of the temporal PSD of the vertical and horizontal components of the velocity as a function of $\omega^*$, for the series of experiments at heating power $P$ increasing from $0$ to $50$~mW. The black solid line $E_z/E_x=2(1-\omega^{*2})/(1+\omega^{*2})$ shows the behavior expected for an ensemble of inertial waves with an axisymmetric distribution of wavevectors. The inset shows the evolution of the mean value of $E_z/E_x$ in the range $\omega^*\in~]1~3]$ as a function of $P$.}
    \label{fig:aniso_compo}
\end{figure}

To assess whether the waves below $\omega^* = 1$ resemble inertial waves, we adopt a robust test that has been already employed in classical fluids~\cite{Campagne2015,Savaro2020,Monsalve2020}: we compute the ratio $E_z/E_x$. As a consequence of the dispersion relation and the transverse nature of inertial waves, this ratio reads $E_z/E_x=2(1-\omega^{*2})/(1+\omega^{*2})$ for an ensemble of waves with axisymmetric distribution of wavevectors (see SM). We plot the observable $E_z/E_x$ in Fig.~\ref{fig:aniso_compo} and compare the experimental data (points) with this theoretical prediction (black line). Despite being very noisy, the data follow the behavior predicted for inertial waves over the frequency range $\omega^*\in[0.3~1]$, independently of the heating power $P$. Moreover, our analysis of the complete dispersion relation for waves propagating in a quantum vortex lattice (see SM) indicates that inertial waves are expected mainly at scales larger than $1.5$~mm and must therefore correspond to the dominant large-scale modes observed in Fig.~\ref{fig:spacetime}, which are indeed oscillating at frequencies in this range.

On the other hand, for waves above $\omega^* = 1$ (focusing on the energetic range of frequency), the observable $E_z/E_x$ appears to be no longer frequency dependent. Specifically, for $\omega^*\in~]1~3]$, its value decreases from about $4$ to $0.7$ as the heating power increases (see the inset of Fig.~\ref{fig:aniso_compo}). This observation is compatible with the scenario of an emergent ensemble of Kelvin waves. This type of waves, restored in He~II by the tension of quantum vortex lines, manifest as high-frequency small-scale oscillations, typically at $\omega^*>1$ and vertical wavelength below $1.5$~mm for our experimental parameters (see SM for details). Furthermore, Kelvin waves are associated with horizontal deformations of the vertically oriented quantum vortices~\cite{Barckicke2026}, which should naturally lead to the reduction in the ratio $E_z/E_x$ as their amplitude increases.

\begin{figure}[tbp!]
    \centering
    \includegraphics{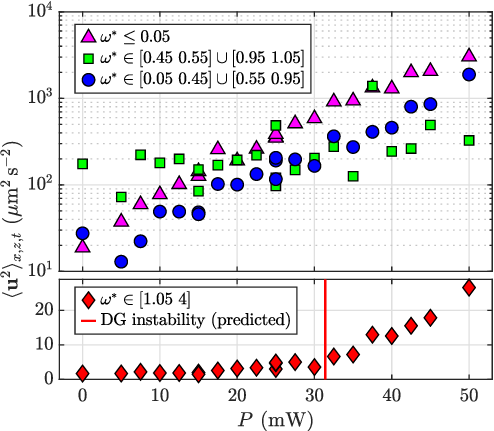}
    \caption{Kinetic energy $\langle {\bf u}^2 \rangle_{x,z,t}$ as a function of the heating power $P$ for the measured velocity field band-pass filtered over four complementary ranges of frequencies (see legend). The series corresponding to $\omega^*\in [1.05~4]$ (red diamonds) is plotted in linear scale to highlight a significant energy increase above $P \approx 30$~mW, which coincides with the onset of the DG instability (red vertical line; see SM for details).}
    \label{fig:ken_vs_P}
\end{figure}

The distinct characteristics of the waves observed above or below the $\omega^* = 1$ limit is further substantiated when we consider how the kinetic energy of these two modes (i.e., the integrated PSD) scales with the heating power $P$. We show this dependence in Fig.~\ref{fig:ken_vs_P} for different frequency bands. First, the energy corresponding to the modes at $\omega^*=0.5$ and $1$ (green squares) appears nearly independent of the heating power $P$, confirming that they result from another forcing mechanism. In parallel, the energy of the mode where inertial wave are dominant (blue circles) follows an approximately exponential increase with $P$, revealing that these inertial waves develop in the system for an arbitrarily small heating amplitude and not as the result of an instability with a finite threshold. The energy of the low frequency mode (magenta triangles)---which also includes the upward drift of the H\textsubscript{2} particles due to the counterflow \cite{LaMantia2016b}---follows a very similar behavior, while the link between these two modes remains to be understood. Unlike these cases, the energy contained in the band $\omega^*\in[1.05~4]$ (red diamonds in the bottom panel) displays a significant increase only above $P\approx 30$~mW, which suggests an alternative mechanism that triggers waves in this frequency domain. This threshold in heating power corresponds well to the value that can be predicted for the already mentioned DG instability, which is $P_{DG} = 31.4$~mW in our system (red vertical line; see SM for details). The DG theoretical instability framework is associated with the growth of inertia-Kelvin waves in the rotating quantum vortex lattice and predicts, in good agreement with our results, the preferential emergence of these waves in a range of frequencies of the order of $\omega^*=2$. This suggests that the waves we observe in the range $\omega^*\in ]1~5]$ are likely to be inertia-Kelvin waves produced by the DG instability.

Finally, over this specific frequency range and above the DG threshold, we observe in Fig.~\ref{fig:PSD} the growth of a continuum of energy which is compatible with a power law. Remarkably, a power-law spectrum in $\omega^{-4/3}$, less steep than the one we can estimate experimentally, is expected for a weak turbulence of Kelvin waves along a quantum vortex~\cite{KW,Lvov2010,Boue2011}. Could the continuous spectrum we observe over the range $\omega^*\in ]1~5]$ be related to such Kelvin-wave turbulence is an open question. At these frequency values, the influence of the inertial-wave dynamics has indeed not completely vanished (see details in SM). This remark must also be tempered by the fact that numerical simulations report strong attenuation of Kelvin waves at the temperature we consider in this work~\cite{Boue2015}. Additional experiments, especially at lower temperature~\cite{Vessaire2026,Golov2026}, are hence required to progress in the understanding of this continuous spectrum arising at $\omega^*>1$ and to investigate whether it could result from the energy cascade of a turbulence of Kelvin waves, of which a recent observation has been achieved in water at room temperature along a single thin-core vortex~\cite{Barckicke2026b}.

\textit{Conclusion.---} We leverage the capacity of small particles to be trapped in the core of quantum vortices to visualize them and investigate, at $T = 2.088$~K, the dynamics of the vortex lattice emerging in rotating He~II. We investigate its weakly perturbed regime, induced by a moderate steady heat flux aligned with the rotation, beyond the space-time averaged indirect diagnoses previously carried out~\cite{Swanson1983}. We reveal that the quantum vortex lattice exhibits a continuous spectrum of propagating waves, whose amplitude scales with the applied heating power, for frequencies up to approximately ten times the rotation rate $\Omega$. Although these waves are expected to follow a dispersion relation that combines three types of excitations (inertial, Kelvin and Tkachenko waves; see SM), we discovered strong signatures of inertial waves at frequencies smaller than $2\Omega$; these modes emerge even at low heating powers. At higher frequencies, we additionally found a threshold in the heating power---in line with the prediction of the Donnelly-Glaberson instability---beyond which a continuous energy spectrum in frequency exhibit premises of a power-law behavior whose slope seems slightly steeper than the $\omega^{-4/3}$ scaling expected for weak turbulence of Kelvin waves along a quantum vortex. Understanding the mechanisms that give rise to these two wave modes---inertial waves at $\omega<2\Omega$ and inertia-Kelvin waves at $\omega>2\Omega$---remains a challenge that could be met by conducting experiments at lower temperatures for which the influence of the normal component of the He~II flow---probably crucial in this separation at $2\Omega$---diminishes.

Indeed, the observed break in temporal PSDs at $2\Omega$ probably exemplifies the impact of mutual friction on finite-temperature vortex dynamics. Given that mutual friction is a temperature-dependent effect~\cite{Donnelly1998}, our description of the weakly perturbed vortex lattice may change with temperature. While exploring these temperature effects motivate immediate experiments, we have demonstrated that visualization studies of perturbed vortex systems in rotating He~II open a new experimental frontier for investigating interactions between quantum vortices and the normal component, in line with state-of-the-art numerical simulations~\cite{Galantucci2020,Galantucci2026,Scollo2026}. These experimental capabilities could prove especially valuable for comparing the spectral content, and, more generally, the coupled three-dimensional quantum vortex dynamics at finite temperature, not only in He~II but across superfluids in general.

\textit{Acknowledgments.---}~This work was supported by the Agence Nationale de la Recherche (grants ANR-23-CE30-0024 and ANR-11-PDOC-0001) and a grant from the Simons Foundation (651461, PPC). We also thank J.I. Polanco and G. Krstulovic for their precious comments.

\clearpage\raggedbottom\newpage

\setcounter{figure}{0}
\setcounter{equation}{0}

\renewcommand\thefigure{S\arabic{figure}}    
\renewcommand\theequation{S\arabic{equation}}

\onecolumngrid

\begin{center}
    {\large \bf ---Supplemental Material---\\
Visualization of Inertial and Kelvin Waves on the Quantum Vortex Lattice in Superfluid Helium}
\end{center}
\bigskip

\section{Waves in a lattice of quantum vortices}

The dispersion relation of waves propagating in a lattice of quantum vortices in a superfluid rotating at a rate $\Omega$ is predicted to be~\cite{smSonin1987}
 \begin{equation}\label{eq:dispersion}
    \omega^2=\left(2\Omega+\nu_s k_z^2\right)\left(2\Omega \frac{k_z^2}{k^2}+\nu_s k_z^2 + \frac{\kappa k_\perp^2}{16\pi} \right)\, ,
 \end{equation}
where $k_z$ is the component along the rotation axis of the wavevector ${\bf k}=(k_x,k_y,k_z)$ of norm $k=|{\bf k}|$ and of horizontal wavenumber $k_\perp=(k_x^2+k_y^2)^{1/2}$. In relation~(\ref{eq:dispersion}), $\nu_s$ denotes the vortex tension and is equal to $\kappa \log(a_m/a_0)/4\pi$ where, for He~II, $\kappa = h/m\simeq 9.97\times10^{-8}~\mathrm{m^2\,s^{-1}}$ is the quantum of circulation, $m$ the mass of the $^4$He atom, and $a_0 \simeq 10^{-10}$~m the radius of the quantum vortex core. The characteristic length $a_m$ is of the order of the intervortex distance $\delta=\sqrt{\kappa/2\Omega}$. This relation~(\ref{eq:dispersion}) pertains to the (discrete) field of the vortex core displacements relative to their equilibrium positions within the lattice at rest. It can be derived by considering the interplay of the velocities induced by the local circulation of each vortex portion on all the portions of the other vortices~\cite{smSonin1987}.

Neglecting the third term of the second parenthesis in relation~(\ref{eq:dispersion}), it is useful to discuss two limits as a function of non-dimensional number $E_s=\nu_s k_z^2/2\Omega$, so noted by analogy with the Ekman number in viscous rotating fluids. First, when $E_s$ is small (i.e., strong rotation, large vertical scale), one recovers the classical dispersion relation $\omega=2\Omega |k_z|/k$ of inertial waves (IW) propagating in rotating fluids~\cite{smMora2021}. This outcome is not surprising, since superfluids are expected to mimic classical fluids at large enough scales. Conversely, for large $E_s$, one finds that $\omega=\nu_s k_z^2$, which corresponds to Kelvin waves (KW) propagating along the cores of quantum~\cite{smBaggaley2014} but also classical vortices~\cite{smBarckicke2026}.

Writing a unified dispersion relation such as Eq.~(\ref{eq:dispersion}) implies investigating scales larger than the intervortex distance~$\delta$. Let us nevertheless note that the Kelvin wave dynamics is expected to persist at scales smaller than $\delta$, which is thought to be crucial for energy dissipation in flows of He~II. Nonlinear mixing of Kelvin waves is indeed proposed to transfer energy to small scales, such turbulent energy cascade ending with phonon emissions when the wavelength approaches the size of the vortex core \cite{smSkrbek2021}. This scenario in particular explains why He~II is expected to remain effectively dissipative even near the absolute zero temperature~\cite{smBarenghi2004}. Coming back to the vortex lattice, at scales below the intervortex distance, the Kelvin wave propagation is driven by the self-induced velocities in a given vortex and no more by the interactions with the surrounding vortex lattice. As a consequence, one has to replace $\delta$ in the expression for the vortex tension by the excitation wavelength, $\lambda_z = 2\pi/k_z$, so that $\nu_s$ becomes $\kappa \log(\lambda_z/a_0)/4\pi$ in the Kelvin wave dispersion relation~\cite{smBoue2011,smBarckicke2026}.

\begin{figure}[]
    \centering
    \includegraphics[width=8cm]{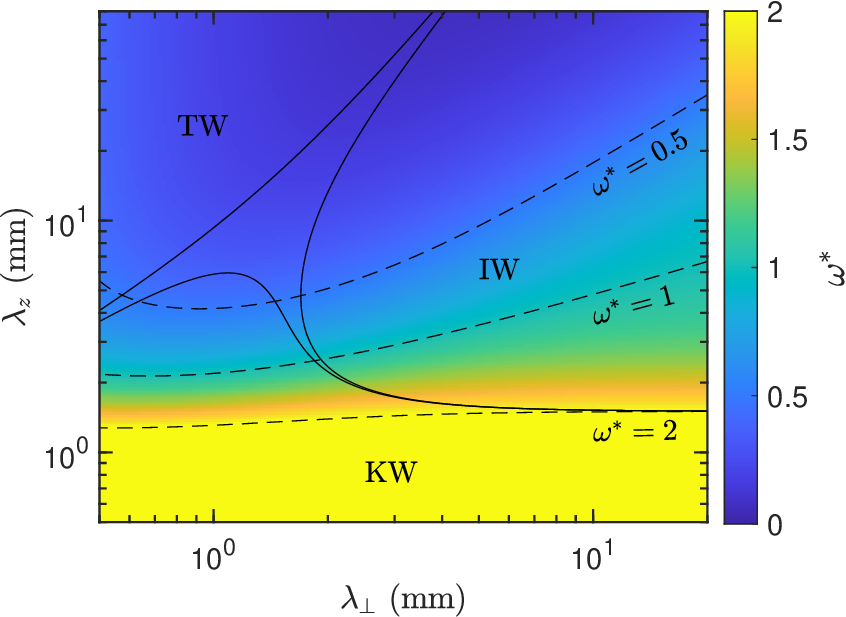}
    \caption{Map of the non-dimensional frequency $\omega^*=\omega/2\Omega$ as a function of the horizontal $\lambda_\perp=2\pi/k_\perp$ and vertical $\lambda_z=2\pi/k_z$ wavelengths for the rotation rate $\Omega=1.047~\mathrm{rad\,s^{-1}}$ used in our experiments. Three black solid lines highlight the limits of the three regions where one term of the second parenthesis in Eq.~(\ref{eq:dispersion}) is larger than the sum of the two other terms, hence identifying regions where the dynamics of inertial waves (IW), Kelvin waves (KW) or Tkachenko waves (TW) dominates. The dashed lines show iso-frequency lines for $\omega^*=0.5, 1$ and $2$.}
    \label{fig:disp}
\end{figure}

To complete this discussion, it is important to consider the full relation~(\ref{eq:dispersion}) including the term $\kappa k_\perp^2/16\pi$ which is associated to a third type of waves, propagating in the plane normal to the rotation axis, called Tkachenko waves (TW)~\cite{smSonin1987,smCoddington2003}. In the experiments reported in this Letter, the rotation rate is $\Omega=1.047~\mathrm{rad\,s^{-1}}$ ($10$~rpm) and results in a lattice mesh of $\delta \simeq 220~\mu$m and a vortex tension $\nu_s \simeq 1.2 \times 10^{-7}~\mathrm{m^2\,s^{-1}}$. Fixing these values allows us to plot in Fig.~\ref{fig:disp} the map of the non-dimensional angular frequency $\omega^*=\omega/2\Omega$ as a function of the horizontal $\lambda_\perp=2\pi/k_\perp$ and vertical $\lambda_z=2\pi/k_z$ wavelengths. From below, the axes are bounded by $500~\mu$m$\,\simeq 2\delta$ and from above, by $\lambda_\perp=2$~cm and $\lambda_z= 8$~cm, which are the dimensions of our experimental domain. In addition to the $\omega^*$ colormap, Fig.~\ref{fig:disp} also highlights the limits of three regions where one term of the second parenthesis in Eq.~(\ref{eq:dispersion}) is larger than the sum of the two other terms: this allows to identify regions in the ($\lambda_\perp$,$\lambda_z$) space where IW, KW or TW dynamics dominates.

Figure~\ref{fig:disp} shows that KW might be discernible in our experimental system for $\lambda_z \lesssim 1.5$~mm (corresponding to $E_s=\nu_s k_z^2/2\Omega \simeq 1$). For larger vertical wavelengths and for $\lambda_\perp \gtrsim 3$~mm, we expect IW to dominate the dynamics. Finally, we notice that TW should occur at large vertical and small horizontal wavelengths. Figure~\ref{fig:disp} also shows that it is difficult to associate a given frequency range to each type of wave dynamics. However, simplifying importantly the picture, we can state that TW dominate the dynamics only for non-dimensional angular frequencies $\omega^*$ below $0.5$. IW might be dominant in the dynamics for modes at frequencies $\omega^*$ below $2$ whereas KW waves are expected to be dominant mainly, but not exclusively, at frequencies $\omega^*$ larger than $2$.

It is important to note that, at finite temperatures, Eq.~(\ref{eq:dispersion}) must be amended in order to account for interactions between quantum vortices and the normal component of He~II. The resulting dispersion relation \cite{smGlaberson1974,smSonin1987,smHenderson2004} is more complex and can even include temperature-dependent corrections~\cite{smScollo2026}.

\section{The experimental thresholds of the transitions observed by Swanson et al.}
\label{suppl:sec:swanson}

In Ref.~\cite{smSwanson1983}, in an experiment similar to the one presented in our Letter, Swanson \textit{et al.} used a second sound gauge to measure the density of quantum vortex lines as a function of the heat flux for several rotation rates $\Omega$ of their rotating cryostat. Working at temperature $1.65$~K, they identified two thresholds in the measured signal, expressed in terms of the expected counterflow velocity $v_{ns}=\dot{q}/\rho_s s\,T$ i.e. the difference between the velocities of the normal and superfluid components, whilst the actual observable was the forcing heat flux $\dot{q}$ ($\rho_s$ is the density of the superfluid component and $s$ the specific entropy of He~II).

Written in terms of the heat flux $\dot{q}$, these thresholds are shown to follow the same scaling, namely $\dot{q}_1 = C_1 \rho_s s\,T\,\sqrt{\Omega}$ and $\dot{q}_2 = C_2 \rho_s s\,T\,\sqrt{\Omega}$, with experimentally determined constants $C_1 = 0.053~\mathrm{cm\,s^{-1/2}}$ and $C_2 = 0.118~\mathrm{cm\,s^{-1/2}}$. Assuming $C_1$ and $C_2$ are indeed constants---e.g., independent of temperature or finite-size effects---we can estimate the corresponding values of the heating power $P$ expected for the two thresholds in our experiments. For our experimental rotation rate $\Omega=1.047~\mathrm{rad\,s^{-1}}$, heating area $A = 4\times10^{-4}~\mathrm{m^2}$, and temperature $T = 2.088$~K, at which $\rho_s = 41.7~\mathrm{kg\,m^{-3}}$ and $s = 1220~\mathrm{J\,K^{-1}\,kg^{-1}}$ \cite{smDonnelly1998}, we find 
\begin{eqnarray}
    P_1= \dot{q}_1 A &=& C_1 \rho_s s\,T\,\sqrt{\Omega}A\simeq 23.1~{\rm mW,~and}\\
    P_2= \dot{q}_2 A &=& C_2 \rho_s s\,T\,\sqrt{\Omega}A\simeq 51.4~{\rm mW.}
\end{eqnarray}
While the second threshold $P_2$ is consistent with the heating power at which we observe the breakdown of the quantum vortex lattice and the superfluid system transitioning into rotating quantum turbulence, the first threshold $P_1$ does not coincide with any noticeable transition in our system. Instead, we find (see Fig.~4 of the Letter, bottom panel) that a significant increase of the energy in the normalized frequency range $\omega/2\Omega \in [1.05~4]$ is observed beyond a heating power $P\simeq 30$~mW corresponding well with the onset referenced in the literature for the Donnelly-Glaberson instability (see next section).

\section{The theoretical threshold of the Donnelly-Glaberson instability}

In an article from 1974~\cite{smGlaberson1974}, Glaberson \textit{et al.} studied theoretically the instability---with respect to inertia-Kelvin waves---of an array of vertical quantum vortices interacting with an upward, uniform and constant flow of the normal fluid component in the framework of the two-fluid model of He~II. Considering first the simplified case of waves propagating in the vertical direction, invariant in the horizontal direction (i.e., with zero horizontal wavenumber) and also neglecting mutual friction in their dispersion relation, they identified a  counterflow velocity threshold for the instability to occur: $v_c = 2\sqrt{2\nu_s \Omega}$. This prediction was shown to describe with ``good qualitative and fair quantitative agreement'' the experimental data of the attenuation of a transverse negative ion beam by Cheng, Cromar, and Donnelly~\cite{smCheng1973} in thermal counterflow of rotating He~II. It is worth to note that, in this theoretical description, the first unstable mode is observed at non-dimensional angular frequency $\omega^*=2$ and for a ``superfluid'' Ekman number $E_s=\nu_s k_z^2/2\Omega=1$ which correspond to the frontier between IW and KW in the wave-type state diagram of Fig.~\ref{fig:disp} (at large $\lambda_\perp$).

Let us now compare the threshold velocity of this instability, which is often called the Donnelly-Glaberson (DG) instability since then, with the counterflow velocity relevant for our experiments. 
In Sec.~\ref{suppl:sec:swanson} of this supplemental material, the relation for the counterflow velocity $v_{ns}=\dot{q}/\rho_s s\,T$ describes the spatially-averaged velocity and does not account for boundary effects relevant in finite-size channels like ours. Following Ref.~\cite{smSwanson1983}, we assume a Poiseuille profile for the normal fluid and a plug profile for the superfluid component. The maximum counterflow velocity at the center of the channel becomes $v_{ns}^{\max} = r v_{ns}$ with the geometric factor $r = 1 + 5\rho_s/4\rho$ for a square channel ($\rho$ is the total density of He~II; at $T = 2.088$~K, $\rho = 146~\mathrm{kg\,m^{-3}}$ and $r = 1.36$ --- see \cite{smPeretti2024} for the derivation). By equating $v_{ns}^{\max}$ to the DG threshold $v_c=2\sqrt{2\nu_s \Omega}$ and rearranging, we obtain the expected threshold heating power for the DG instability,
\begin{equation}
    P_{DG} = 2 A \rho_s s T\, \frac{\sqrt{2\nu_s\Omega}}{r} \simeq 31.4~{\rm mW}.
\end{equation}
This value is indicated by the red vertical line in the bottom panel of Fig.~4 in the main text.

Similarly, the DG threshold has been invoked by Swanson \textit{et al.}~\cite{smSwanson1983} (see previous section) to explain their first transition experimentally observed for a heat flux $\dot{q}_1 = C_1 \rho_s s\,T\,\sqrt{\Omega}$ with $C_1 = 0.053~\mathrm{cm\,s^{-1/2}}$ and therefore for the maximum counterflow velocity $v_{ns}^{\max} = r C_1\,\sqrt{\Omega}$ at the center of their experimental channel (in their experiments, $T = 1.65$~K and therefore $r = 2.01$). The experimentally measured constant $C_1$ was, introducing the Poiseuille correction $r$, shown to quantitatively correspond to the value of the theoretical prediction $2\sqrt{2\nu_s}$ by Glaberson \textit{et al.}

Nevertheless, the DG instability is shown to be more complex in the theoretical paper by Glaberson \textit{et al.}~\cite{smGlaberson1974}. Indeed, allowing instabilities with respect to waves propagating in tilted directions, the predicted threshold of the instability is shown to decrease, possibly down to $v_c/2 = \sqrt{2\nu_s \Omega}$. Considering mutual friction with the normal fluid component will also surely modify the predicted threshold for the instability of an array of quantum vortices in interaction with a parallel flow of normal fluid. We leave this theoretical subject aside for future works since it is out of the scope of our Letter to quantitatively explain the threshold in heating power we observe in the experiments we report. We should nevertheless highlight that the very quantitative agreement reported by Swanson \textit{et al.}~\cite{smSwanson1983} between their first observed transition and the DG prediction for the constant $C_1$ might actually only be a qualitative agreement. 

\section{Space-time diagrams of the velocity field for the modes at $\omega^*=0.5$ and $1$}

\begin{figure}[h!]
    \centering
    \includegraphics[scale=1]{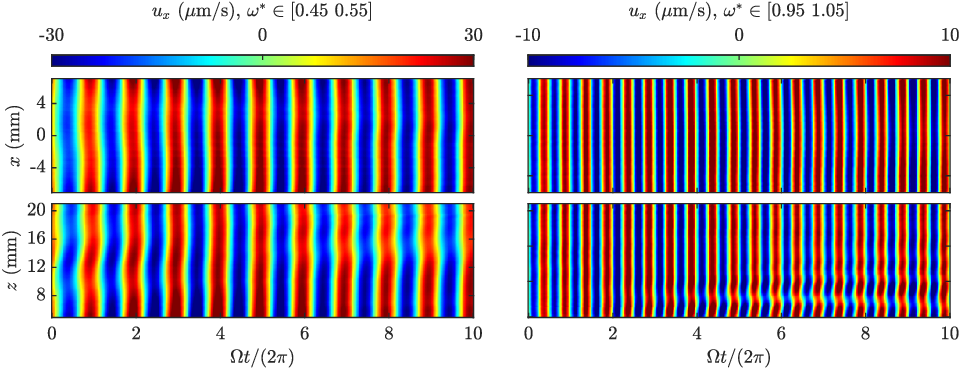}
    \caption{Space-time diagrams of the horizontal velocity for an experiment at $P = 5$~mW, frequency-filtered in the range $\omega^* \in [0.45~0.55]$ (left column) and $\omega^* \in [0.95~1.05]$ (right column), for $z = 10$~mm (top row) and $x = 0$~mm (bottom row). Time $t$ is normalized by the cryostat's period of rotation $2\pi/\Omega$. $t=0$ indicates the moment when the heater is switched on, $x=0$ the horizontal center of the He~II channel and $z=0$ the vertical position of the heater surface.}
    \label{fig:spacetime5mW}
\end{figure}

In Fig.~\ref{fig:spacetime5mW}, we report the space-time diagrams of the horizontal velocity for an experiment at $P = 5$~mW, frequency-filtered in the range $\omega^* \in [0.45~0.55]$ (left column) and $\omega^* \in [0.95~1.05]$ (right column), for $z = 10$~mm (top row) and $x = 0$~mm (bottom row). In $x$-axis, time $t$ is normalized by the cryostat's period of rotation $2\pi/\Omega$. $t=0$ indicates the moment when the heater is switched on, $x=0$ the horizontal center of the He~II channel and $z=0$ the vertical position of the heater surface.

These figures show that the modes at $\omega^*=0.5$ and $1$ that we detect in our experiments are associated to large-scale standing modes both in the horizontal and vertical directions. These modes are actually nearly homogeneous over the region of the vertical $(x,z)$ plane where the velocity field is measured.

\section{Componential anisotropy of a statistically axisymmetric ensemble of inertial waves}

Inertial waves are the solutions of the linear limit of the Navier-Stokes equation in a rotating frame
\begin{eqnarray}
    \frac{\partial {\bf u}}{\partial t}&=&-\frac{1}{\rho}\nabla p -2{\mathbf \Omega}\times{\bf u}\, ,
\end{eqnarray}
where we also took the inviscid limit for the sake of simplicity. Combining this equation with the incompressibility condition and injecting a plane wave in the equations, one gets the dispersion relation of inertial waves
\begin{eqnarray}
    \omega=\hat{s} 2\Omega\frac{k_z}{k}\, ,
\end{eqnarray}
where $k$ is the norm of the wavevector ${\bf k}=(k_x,k_y,k_z)$, $\hat{s}=\pm 1$ and $\omega>0$.

Let us write the velocity field of a plane inertial wave of wavevector ${\bf k}=(k_X,0,k_Z)$ and amplitude $A$. We choose here the cartesian coordinate system $(X,Y,Z)$ such that $k_X>0$, $k_Y=0$ and $Z$ is oriented along the rotation axis, without loss of generality. The helical velocity field of the wave writes
\begin{eqnarray}
    u_X &=& - \hat{s} \omega^* k A \sin({\bf k}\cdot {\bf X}-\omega t)\, ,\\
    u_Y &=& \hat{s} k A \cos({\bf k}\cdot {\bf X}-\omega t)\, ,\\
    u_Z &=& \sqrt{1-{\omega^*}^2}k A \sin({\bf k}\cdot {\bf X}-\omega t)\, ,
\end{eqnarray}
which expressions can be obtained for example from Ref.~\cite{smBordes2012} using the relations $k_Z=\hat{s} \omega^* k$ and $k_X=\sqrt{1-{\omega^*}^2}k$ with $\omega^*=\omega/2\Omega$.

We now introduce a second cartesian coordinate system $(x,y,z)$ with $x=X\cos{\varphi}-Y\sin{\varphi}$, $y=X\sin{\varphi}+Y\cos{\varphi}$ and $z=Z$. The plane $(x,z)$ can for example be the PIV measurement plane which is, in the general case, not the plane in which the wave propagates. Placing ourselves at ${\bf X}={\bf 0}$, without loss of generality, we have
\begin{eqnarray}
    u_x &=& u_X\cos{\varphi}-u_Y\sin{\varphi}= \hat{s} A k \left(\omega^* \sin(\omega t)\cos{\varphi}- \cos(\omega t)\sin{\varphi}\right)\, ,\label{eq:ux}\\
    u_y &=& u_X\sin{\varphi}+u_Y\cos{\varphi}= \hat{s} A k \left(\omega^* \sin(\omega t)\sin{\varphi}+ \cos(\omega t)\cos{\varphi}\right)\, ,\label{eq:uy}\\
    u_z &=& u_Z = -\sqrt{1-{\omega^*}^2}k A \sin(\omega t)\, .\label{eq:uz}
\end{eqnarray}

Let us now assume that we are looking at a flow composed of an ensemble of statistically independent inertial waves with a random phase distribution, the same wavelength, amplitude and frequency but with an axisymetric distribution of wavevector direction around the rotation axis in the Fourier space. Using Eqs.~(\ref{eq:ux}) and (\ref{eq:uz}), we can first explicit the components $u_x$ and $u_z$ of such a velocity field as
\begin{eqnarray}
    u_x &=& \frac{1}{N} k A \sum_{n=1}^N \hat{s}_n  \left[\omega^* \sin(\omega t+\phi_n)\cos{\varphi_n} - \cos(\omega t+\phi_n)\sin{\varphi_n} \right] \, , \\
    u_z &=& -\frac{1}{N} \sqrt{1-{\omega^*}^2} k A \sum_{n=1}^N \sin(\omega t+\phi_n) \, .
\end{eqnarray}
Taking the temporal average of the associated kinetic energies further gives (cancelling the sum of the cross terms which vanish on average)
\begin{eqnarray}
    \left \langle u_x^2 \right \rangle_t  &=& \frac{1}{2N} k^2 A^2 \sum_{n=1}^N  \left[{\omega^*}^2 \cos^2{\varphi_n} + \sin^2{\varphi_n} \right] \, , \\
    \left \langle u_z^2 \right \rangle_t  &=& \frac{1}{2N} (1-{\omega^*}^2) k^2 A^2 N \, .
\end{eqnarray}
The distribution of the wavevector horizontal direction $\varphi_n$ being also uniform, we obtain (with $N\rightarrow \infty$)
\begin{eqnarray}
    \left\langle u_x^2  \right\rangle_t &=& \frac{1}{4} k^2 A^2 ({\omega^*}^2 + 1)   \, ,\\
    \left\langle u_z^2  \right\rangle_t &=& \frac{1}{2} (1-{\omega^*}^2) k^2 A^2 \, ,
\end{eqnarray}
and we finally have
\begin{eqnarray}
    \frac{\left\langle u_z^2  \right\rangle_t}{\left\langle u_x^2  \right\rangle_t} &=& 2\frac{1-{\omega^*}^2 }{1+{\omega^*}^2 }\, .
\end{eqnarray}
This result naturally extends to the ratio of the temporal power spectral densities of the vertical and horizontal velocity components that we present in the Letter.

\end{document}